\documentclass[pra,twocolumn,amsmath,nofootinbib,longbibliography]{revtex4-1}

\usepackage[bookmarks = false, pdfpagemode = None, pdfstartview = FitH, colorlinks = true, urlcolor = blue,hyperfootnotes=false]{hyperref}
\usepackage{rotating}

\newcommand{\ket}[1]{{\left\vert{#1}\right\rangle}}

\newcommand{\C}[1]{{{\vphantom{#1}}^{C}\hspace{-.3em}{#1}}}
\newcommand{\CC}[1]{{{\vphantom{#1}}^{CC}\hspace{-.3em}{#1}}}

\newcommand{\CZ}{\C{Z}}
\newcommand{\CCZ}{\CC{Z}}

\newcommand{\e}{\textrm{e}}
\renewcommand{\i}{\textrm{i}}

\begin{document}

\title{Distilling one-qubit magic states into {T}offoli states}

\author{Bryan Eastin}
\email[]{Bryan.Eastin@ngc.com}
\affiliation{Northrop Grumman Corporation, Baltimore, MD}

\date{\today}

\begin{abstract}
For certain quantum architectures and algorithms, most of the required resources are consumed during the distillation of one-qubit magic states for use in performing Toffoli gates.  I show that the overhead for magic-state distillation can be reduced by merging distillation with the implementation of Toffoli gates.  The resulting routine distills $8$ one-qubit magic states directly to a Toffoli state, which can be used without further magic to perform a Toffoli gate.
\end{abstract}


\maketitle

Quantum algorithms frequently include a reversible classical subroutine that dominates the computation.  Consequently, the Toffoli gate, which is universal for classical reversible computing, is commonly the most-used gate in an algorithm.  Toffoli gates are inconvenient in many quantum architectures, but they can be implemented using, for example, one-qubit magic states and Clifford gates, where the Clifford gates are taken here to include both unitary Clifford operators and measurement and preparation in Pauli eigenbases.  The initial preparation of magic states is generally poor, so prior to use, it is necessary to distill them, increasing their fidelity with the intended state.  Magic-state distillation can be a significant burden.  Recent quantum architecture papers indicate that when running Shor's algorithm on interesting problem sizes $90\%$ of the physical qubits can easily be devoted to magic-state distillation~\cite{Jones12,Fowler12b}.  Such observations have helped to spur a significant body of new work focused on reducing the resources required by magic-state distillation routines~\cite{Meier12,Anderson12,Jochym-O'Connor12,Campbell12,Bravyi12,Jones12b}.

With a few exceptions~\cite{Reichardt05,Reichardt06c,Campbell11}, past research on magic-state distillation has focused on routines that transform multiple faulty copies of a magic state into fewer improved copies of the same state.  Being simple to inject, one-qubit magic states are a natural starting point for distillation, and routines that output the same sort of state as the input are convenient.  Ultimately, however, the reason for distilling magic states is frequently to implement a Toffoli gate.  Rather than segregating the two tasks, I show in this paper that one can combine them to obtain reductions in the resources required to implement a Toffoli gate.

I describe here a novel magic-state distillation routine, the $H$-to-Toffoli routine, that takes $8$ copies of the one-qubit magic state $\ket{H}=\cos(\frac{\pi}{8})\ket{0} + \sin(\frac{\pi}{8})\ket{1}$ that suffer $Y$ errors with probability $p$ and, on success, outputs a single Toffoli state, $\ket{\text{Toffoli}}$, that suffers errors with a probability of roughly $28 p^2$.  One measure of the efficiency of a distillation routine is the state cost, that is, the number of input copies of the magic state required per improved output.  Using the $H$-to-Toffoli routine, the state cost for implementing a Toffoli gate with quadratically reduced error is competitive with the most efficient distillation routines known~\cite{Bravyi12,Jones12b}.  Moreover, the location cost, which I define as the number of locations in the distillation circuit per output, is smaller by a factor of $2$ to $4$ for the same task.

Aside from the aforementioned differences, the scenario considered here is the standard one for magic state distillation~\cite{Bravyi05}: Clifford gates are taken to be perfect while magic states are assumed to suffer from a limited (by twirling) set of errors.  The notation largely follows that of Meier et al.~\cite{Meier12}.

The remainder of this paper is organized as follows: Section~\ref{sec:HToToffoliDistillation}
introduces the $H$-to-Toffoli routine, and Sec.~\ref{sec:ToffoliDistillation} explains how a related approach can be used for distilling Toffoli states.  The efficiency of the $H$-to-Toffoli routine and its relative performance are discussed in Sec.~\ref{sec:efficiency}.  The conclusion appears in Sec.~\ref{sec:conclusion}, and the circuits used in the calculation of location costs are given in the appendix.

\begin{figure}
\includegraphics[clip = true, trim = .2cm 0cm 0cm 0cm]{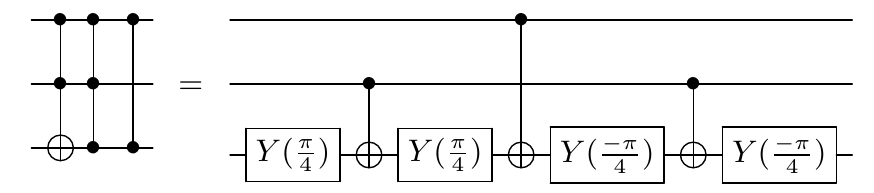}
\caption{Decompositions of the Margolus-Toffoli gate.  (See Ref.~\cite{Song04}.)  On the left is a decomposition of the Margolus-Toffoli gate in terms of a true Toffoli gate, a controlled-controlled-sign ($\protect\CCZ$) gate, and a controlled-sign ($\protect\CZ$) gate.  Note that the Margolus-Toffoli gate is equivalent to a Toffoli gate followed (or preceded) by the transformation $\ket{101}\rightarrow -\ket{101}$.  On the right is a decomposition in terms of Clifford gates and $\pi/4$ rotations about the $Y$ axis of the Bloch sphere. \label{fig:MargolusToffoliDecompositions}}
\end{figure}

\section{$H$-to-Toffoli distillation\label{sec:HToToffoliDistillation}}

\begin{figure}
\includegraphics{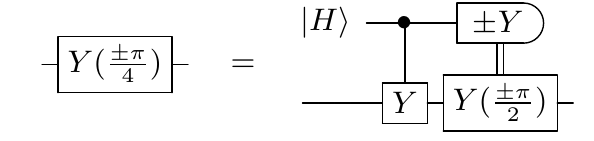}
\caption{Circuit identity showing how the rotation $Y(\frac{\pm\pi}{4}) = \cos(\frac{\pi}{8})I \mp \i \sin(\frac{\pi}{8})Y$ can be implemented using Clifford gates, the state $\ket{H}$, and the (Clifford) rotation $Y(\frac{\pm\pi}{2})$. \label{fig:indirectYrotation}}
\end{figure}

\begin{figure}
\includegraphics{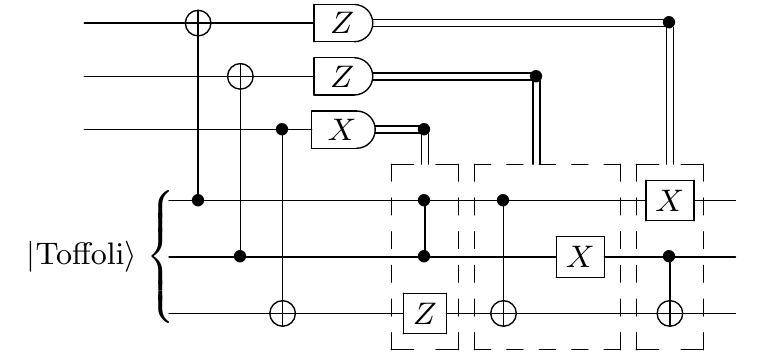}
\caption{A circuit implementing the Toffoli gate using a Toffoli state, $\ket{\text{Toffoli}}=(\ket{000}+\ket{100}+\ket{010}+\ket{111})/2$, as per Shor~\cite{Shor96}. The target of the Toffoli gate corresponds to the third qubit in each block. \label{fig:indirectToffoli}}
\end{figure}

At the heart of the new distillation routine are two observations: First, the standard circuit for the Margolus-Toffoli gate (shown in Fig.~\ref{fig:MargolusToffoliDecompositions}), when implemented using twirled faulty $\ket{H}$ states (see Fig.~\ref{fig:indirectYrotation}), is equivalent to a perfect Margolous-Toffoli gate potentially followed by $Y$ errors on the target and $Z$ errors on the controls, and furthermore, an error on a single $\ket{H}$ state always results in a $Y$ error on the target.  Second, given Margolus-Toffoli gates that occasionally suffer from $Y$ (or $X$) errors on the target one can use several such gates to prepare a Toffoli state with multiple target qubits and then check the target qubits against each other to reduce the probability of an undetected error on the prepared Toffoli state.  Verified Toffoli states can then be used to implement Toffoli gates using the indirect method of Shor (shown in Fig.~\ref{fig:indirectToffoli}), or they can be further checked against each other.

\begin{figure*}
\includegraphics{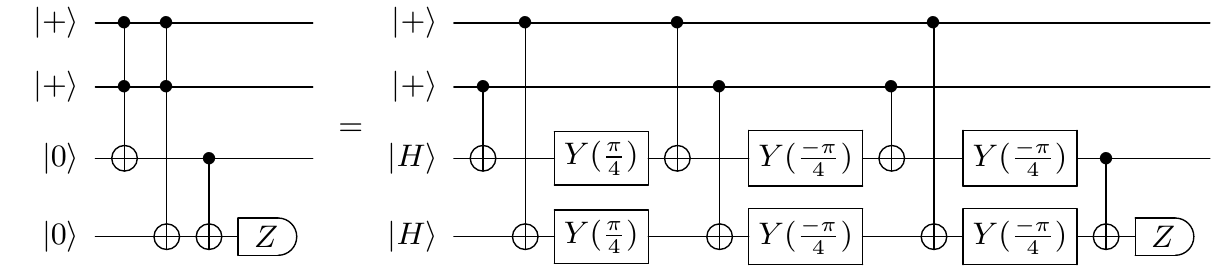}
\caption{$H$-to-Toffoli distillation circuit. The output is discarded whenever a non-trivial measurement outcome is obtained.  The circuit on the left shows the distillation in terms of Toffoli gates while the circuit on the right shows the same distillation circuit expanded in terms of the $\ket{H}$-state implementation of Margolus-Toffoli gates.  All $Y(\frac{\pm\pi}{4})$ gates are implemented indirectly as in Fig.~\ref{fig:indirectYrotation}.  By enumeration and error propagation it is easily shown that, to lowest non-trivial order, this circuit takes $\ket{H}$ states that suffer $Y$ errors with probability $p$ to Toffoli states that suffer errors (some combination of $X$ errors on the target qubit and $Z$ errors on the controls) with probability $28 p^2$. \label{fig:HToToffoliDistillation}}
\end{figure*}

\begin{figure}
\includegraphics{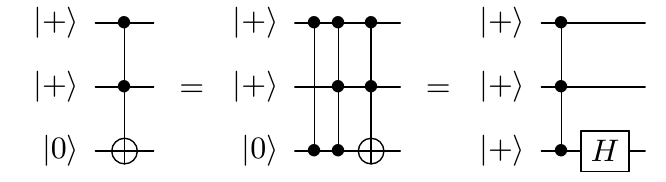}
\caption{Toffoli state preparation. The circuit on the left shows an obvious method of preparing the Toffoli state.  The middle circuit shows Toffoli-state preparation using the Margolus-Toffoli gate (decomposed into three more familiar gates).  From the right circuit it is clear that the target qubit of a Toffoli state can be changed using a pair of Hadamard gates.  The first equality follows from the fact that a gate controlled on $\ket{0}$ is not executed; the same logic implies that a Margolus-Toffoli can be substituted for a true Toffoli gate whenever the target qubit is initially prepared in the state $\ket{0}$, as will be the case whenever the Margolus-Toffoli gate is used in the distillation routines presented here. \label{fig:ToffoliStatePreparation}}
\end{figure}

The $H$-to-Toffoli magic-state distillation routine is shown in Fig.~\ref{fig:HToToffoliDistillation}.  Both Margolus-Toffoli gates use the same control qubits, but they have different targets.  On such input the Margolus-Toffoli gate acts like  the Toffoli gate (see Fig.~\ref{fig:ToffoliStatePreparation}), a classical reversible gate.  Consequently, in the absence of errors, measuring the two target qubits in the $Z$-eigenbasis would yield the same result, and thus the parity measurement will yield $0$.  It is straightforward to show that an error on an $\ket{H}$ state used to implement the Margolus-Toffoli gate as shown in Figs.~\ref{fig:MargolusToffoliDecompositions} and~\ref{fig:indirectYrotation} can be propagated to a $Y$ error on the target qubit together with, possibly, $Z$ errors on the control qubits.  A single such error will thus be detected by the parity measurement while any two errors will go undetected.  To lowest non-trivial order the acceptance probability $a(p)$ is thus $1-8 p$ and the output error probability $e(p)$ is $\binom{8}{2} p^2 = 28 p^2$.  Exhaustive counting yields
\begin{align*}
 a(p) = &1-8 p + 56 p^2 - 224 p^3 + 560 p^4 \\
&- 896 p^5 + 896 p^6 - 512 p^7 + 128 p^8 \text{\hspace{3em}and}\\
 e(p)a(p) = &28 p^2 - 168 p^3 + 476 p^4 \\
&- 784 p^5 + 784 p^6 - 448 p^7 + 112 p^8\;.
\end{align*}
Conditional on acceptance, only $Z$ errors afflict the output control qubits, while the output target qubit is afflicted only by $X$ errors.  As it happens, each of the seven possible non-trivial errors is equally likely.

The probability of an undetected $X$ error on the output target qubit can be made arbitrarily small by generating more target qubits and checking their parities; the $X$-error probability can be reduced to $O(p^o)$ by generating $o$ target qubits and checking them against one another.  This does not reduce the probability of a $Z$ error on the output control qubits below $O(p^2)$; in fact, the coefficient of $p^2$ worsens as $o$ becomes larger.  If further reductions in the probability of error on the control qubits are necessary, one can resort to generic Toffoli-state distillation.

\section{Toffoli-state distillation\label{sec:ToffoliDistillation}}

For Toffoli-state distillation, it is helpful to restrict the errors that must be considered to $Z$ errors on the control qubits and $X$ errors on the target qubit.  Conveniently, the $H$-to-Toffoli routine outputs states with errors of just this form.  Nevertheless, should it be necessary, Aliferis has shown that the desired error model can be enforced by twirling the Toffoli state with the appropriate set of Clifford gates~\cite{Aliferis07c}.

Toffoli states that suffer only $X$ errors on the target qubit and $Z$ errors on the control qubits can be used to implement Toffoli gates that suffer only $X$ errors on the target and $Z$ errors on the (matching) controls.  Consequently, a quadratic reduction in the probability of an error on the target qubit of such a state can be achieved by implementing the left circuit of Fig.~\ref{fig:HToToffoliDistillation} using Toffoli states.  This distillation routine can be shown to be equivalent to the Toffoli-state distillation routine proposed by Aliferis~\cite{Aliferis07c}.  Given identically prepared Toffoli states, the probability of an $X$ error on the target is reduced from $p$ to roughly $p^2$ and the probability of errors that do not involve an $X$ error on the target roughly doubles.
Reduction of the $Z$-error probability on a control qubit can be achieved using the same circuit if one first swaps the target qubit with the control qubit of the Toffoli state using a pair of Hadamard gates (see Fig.~\ref{fig:ToffoliStatePreparation}).  This transformation takes $Z$ errors on the former control qubit to $X$ errors on the new target qubit and vice versa.

\section{Efficiency\label{sec:efficiency}}

The $H$-to-Toffoli routine is atypical of magic-state distillation routines in that its inputs and outputs are different and, as a consequence, it is not composable with itself.  This complicates comparisons with other distillation routines.  Taking the Toffoli state to have a ``value'' of 4 $\ket{H}$ states, it might be said that the $H$-to-Toffoli routine costs 2 input $\ket{H}$ states per output $\ket{H}$ state with quadratically reduced error probability, numbers which correspond to a scaling exponent of $\log_2 2 = 1$.  This would make it competitive with the most efficient distillation routines~\cite{Bravyi12,Jones12b}, but the comparison is unfair in two ways: The output is not really a collection of $\ket{H}$ states, and the routine is not scalable, being fixed in size and non-composable.  For this reason, I consider below the overhead required for specific tasks: the production of a Toffoli state or Toffoli gate using faulty $\ket{H}$ states, where the ($\ket{H}$-state) inputs and (state or gate) outputs suffer errors with probability $p$ and $O(p^2)$, respectively.

\begin{figure}
\includegraphics{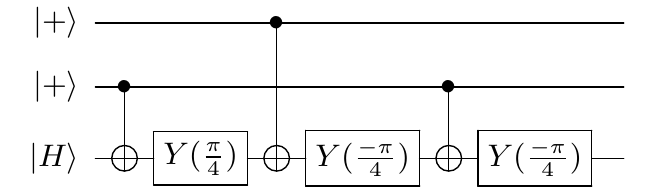}
\caption{Toffoli-state preparation circuit based on the Margolus-Toffoli gate.  This circuit can be used, together with circuits from Figs.~\ref{fig:indirectToffoli} and~\ref{fig:indirectYrotation}, to implement a Toffoli gate using only $4$ $\ket{H}$ states, as opposed to $7$ $\ket{H}$ states as commonly assumed in the literature.  \label{fig:ToffoliStatePreparationExpanded}}
\end{figure}

Using the $H$-to-Toffoli routine, $8$ $\ket{H}$-type magic states which suffer $Y$ errors with probability $p$ are required to distill a single Toffoli state which suffers errors with probability $O(p^2)$.  As illustrated in Fig.~\ref{fig:indirectToffoli}, Clifford gates and a Toffoli state suffice to perform a Toffoli gate, so a Toffoli gate can be implemented with the same parameters.  I assume in this analysis that all Toffoli gates are performed using Toffoli states.  For distillation routines other than the $H$-to-Toffoli routine, $\ket{H}$ states are distilled prior to being used in the Toffoli-state preparation circuit shown in Fig.~\ref{fig:ToffoliStatePreparationExpanded}.  Using this circuit, only $4$ $\ket{H}$ states are required to prepare each Toffoli state and therefore to implement each Toffoli gate.  Bravyi and Haah have shown that $\ket{H}$ states with quadratically suppressed errors can be prepared at a cost arbitrarily close to $3$ input $\ket{H}$ states per output $\ket{H}$ state~\cite{Bravyi12}.  Jones has further shown that quadratic error suppression can be obtained at a state cost arbitrarily close to $2$ as part of a larger distillation routine~\cite{Jones12b}.  Multiplying each of these numbers by $4$, one finds that the $H$-to-Toffoli routine yields an improvement of $33\%$ in the state cost compared to the best routines of Bravyi and Haah and performs similarly to Jones' routines.

\begin{figure}
\includegraphics{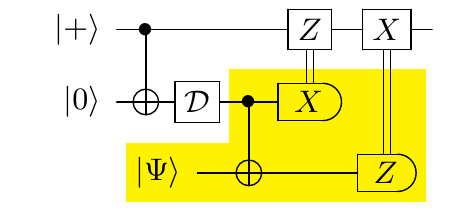}
\caption{(Color online) State injection.  The circuit shown injects an arbitrary state $\ket{\Psi}$ into a quantum code.  The unshaded portion of the circuit is implemented on encoded qubits, while the shaded gates are performed on unencoded qubits.  The gate $\mathcal{D}$ represents decoding the quantum code. \label{fig:stateInjection}}
\end{figure}

\begin{table*}
\begin{tabular}{c@{\hspace{2em}}c@{\hspace{2em}}c@{\hspace{2em}}c@{\hspace{2em}}c}
& & & \multicolumn{2}{c}{Location cost} \\
\cline{4-5}
Distillation routine & State cost & Output error probability & Toffoli state & Toffoli gate\\
\hline \hline
$10$-to-$2$ & $20$ & $36 p^2$ & $183$ & $298$ \\
\hline
$14$-to-$2$ & $28$ & $28 p^2$ & $179$ & $334$ \\
\hline
$26$-to-$6$ & $17.33$ & $76 p^2$ & $151$ & $252.7$ \\
\hline
$H$-to-Toffoli & $8$ & $28 p^2$ & $36$ & $91$
\end{tabular}
\caption{Properties of Toffoli-gate and Toffoli-state implementations based on a selection of distillation routines.  The $14$-to-$2$ and $26$-to-$6$ routines are members of a family of routines developed by Bravyi and Haah~\cite{Bravyi12}.  Within that family the $26$-to-$6$ routine seems to have the lowest location cost.  The $10$-to-$2$ routine is a relatively efficient routine proposed in Ref.~\cite{Meier12}.  The circuits on which the location counts for these distillation routines are based appear in the appendix.  The location costs quoted for the Toffoli gate include both the circuit in Fig.~\ref{fig:indirectToffoli} and the cost of state injection.   \label{tab:HToToffoliRoutineStats}}
\end{table*}

In addition to the state cost, I calculate location costs for the $H$-to-Toffoli distillation routine and some close competitors.  Locations are simply points in (discretized) space and time where a qubit is undergoing a gate or storing quantum information.  The location cost of a distillation routine is the number of locations required per output.
In an effort to make the cost less dependent on the native gate set, one-qubit unitary Clifford gates are ignored when counting locations; one-qubit Clifford measurements are also ignored on the grounds that these tend to be much faster than two-qubit Clifford gates on robustly encoded qubits.

As illustrated by Fig.~\ref{fig:overheadHToToffoliDistillation} in the appendix, the number of locations in the $H$-to-Toffoli distillation routine is $36$.  The number of additional locations required to implement the Toffoli gate as shown in Fig.~\ref{fig:indirectToffoli} using the resultant Toffoli state is $15$.  The total resource cost for implementing a Toffoli gate is thus $51$ locations and $8$ $\ket{H}$ states.  If each $\ket{H}$ state is injected from a lower level of encoding using the circuit in Fig.~\ref{fig:stateInjection} then $5$ additional locations are required per $\ket{H}$ state and the total location cost rises to $91$ locations per Toffoli gate.  By comparison, the appendix contains a compacted circuit for the $26$-to-$6$ routine that requires only $53.7$ locations per improved $\ket{H}$ state counting state injection and $32$ discounting it.  The circuit in Fig.~\ref{fig:ToffoliStatePreparationExpanded} requires $23$ locations and $4$ $\ket{H}$ states.  Thus, using the $26$-to-$6$ routine, $151$ locations are required to produce a Toffoli state with quadratically reduced error and $252.7$, including state injection, for a similarly improved Toffoli gate.  The $H$-to-Toffoli routine reduces the location cost for Toffoli states and gates by a factor of $4.2$ and $2.8$, respectively.  State injection is included in the latter number as an indication of the true cost of a Toffoli gate.  This represents the lower end of the possible location overhead per input $\ket{H}$ state; $\ket{H}$ states having previously undergone distillation will cost many more locations, thereby enhancing the relative performance of the $H$-to-Toffoli routine.

The properties of Toffoli-state and Toffoli-gate implementations using some of the most resource-efficient distillation routines are summarized in Tab.~\ref{tab:HToToffoliRoutineStats}.

\section{Conclusion\label{sec:conclusion}}

Magic-state distillation, as required for the implementation of robust Toffoli gates, is a significant driver of overhead in some proposed quantum computing architectures.
In this paper I have shown that the overhead can be reduced by merging magic-state distillation with the implementation of Toffoli gates.  I described a novel distillation routine that distills $8$ copies of a one-qubit magic state, $\ket{H}$, directly into a Toffoli state with quadratically reduced error probability.  For the purpose of implementing Toffoli gates, the state cost of this routine is as good as or better than existing routines and the location cost is a factor of $2$ to $4$ lower compared to other distillation routines that provide quadratic error suppression.

Subsequent to the development of the $H$-to-Toffoli routine described herein, I learned that Cody Jones has developed a closely related routine, albeit one with a distinct motivation and a very different looking quantum circuit.  Jones' routine likewise distills $8$ copies of a one-qubit magic state into a single Toffoli state and provides the same amount of error suppression.  More information on this complementary work can be found in Ref.~\cite{Jones12c}.

My analysis has focused on a single round of magic-state distillation.  Partly this reflects the fact that the proposed routine outputs Toffoli states, for which few distillation routines are known, and partly it reflects a belief that many rounds of distillation at some fixed level of encoding will be uncommon, as has previously been argued in Refs.~\cite{Raussendorf07,Bravyi12}.  Distillation routines that provide a quadratic reduction in error per round are among the most efficient~\cite{Bravyi12}, and it is unnecessary to perform distillation using encoded gates that are significantly less error prone that the outputs of the distillation routine~\cite{Raussendorf07,Bravyi12,Jochym-O'Connor12}, so it seems likely that maximum efficiency will typically be achieved by doing magic-state distillation at many different levels of encoding, where a quadratic reduction in the error probability is achieved at each level.  As illustrated in the recent architectural paper of Fowler et al.~\cite{Fowler12b}, surface codes are well suited to this approach, being very flexible in the degree of encoding.  For concatenated coding, the steps in the degree of encoding will typically be greater, so routines with stronger error suppression may play a role, but given the difference in the sizes of the coefficients for the output error probabilities for magic-state distillation as opposed to fault-tolerant error correction, I would not expect very high-order distillation to be necessary.  

I have introduced the notion of location cost because I feel that this more accurately identifies the resource that one wishes to minimize in order to control the overhead for fault-tolerant quantum computing.
In the limit of many rounds of distillation, the state cost becomes the determining factor in the location cost, but when only a few rounds of distillation are required the two can yield significantly different rankings of distillation routines.  The location cost is not a particularly portable measure; the best method of counting locations will vary between different architectures, as is illustrated by a recent paper optimizing the location cost of the $15$-to-$1$ magic-state distillation routine for a surface code architecture~\cite{Fowler12c}.  Additionally, while I have focused on the implementation of the Toffoli gate, it is perhaps worth mentioning that the Margolus-Toffoli gate can be implemented at a slightly lower location cost for some of the distillation routines considered here using the circuit in Fig~\ref{fig:MargolusToffoliDecompositions}.  As has been previously noted, the Margolus-Toffoli gate is often an acceptable stand-in for the Toffoli gate~\cite{Barenco95}; in particular, the Margolus-Toffoli gate can be used in a quantum computation to calculate any classical function that is subsequently exactly undone, an occurrence not uncommon in reversible computing.

On the topic of magic-state distillation, several interesting open problems remain.  Thus far, Toffoli-state distillation has received little attention.  The only published routine has a state cost of $8$ for quadratic error suppression~\cite{Aliferis07c}, though a more efficient method was recently suggested by Jones in the conclusion of Ref.~\cite{Jones12}. 
An obvious avenue of investigation is thus to search for new routines distilling Toffoli states.  The compaction of existing distillation routines to minimize their location costs in various architectural settings is another worthwhile endeavor.   
Aside from magic-state distillation, other (older) techniques exist for implementing non-Clifford gates in concatenated codes~\cite{Shor96} and there even exist quantum codes, including some varieties of surface code, that require no distillation at all~\cite{Knill96b,Bombin07}.  
One might therefore investigate the performance tradeoff between magic-state distillation and these other techniques for implementing fault-tolerant non-Clifford gates.  Finally, despite the caveats listed above, it would be interesting to determine the location cost as a function of the target error rate for a variety of distillation sequences under the assumption of perfect Clifford gates.

\acknowledgments

I am grateful to Adam Meier, Manny Knill, Ben Reichardt, Kevin Obenland, and Andrew Landahl for enlightening discussions on this topic.  Special thanks go to Cody Jones for pointing me to other routines for the distillation of Toffoli states.  The magic-state distillation routines described herein are patent pending.

\begin{figure*}
\includegraphics{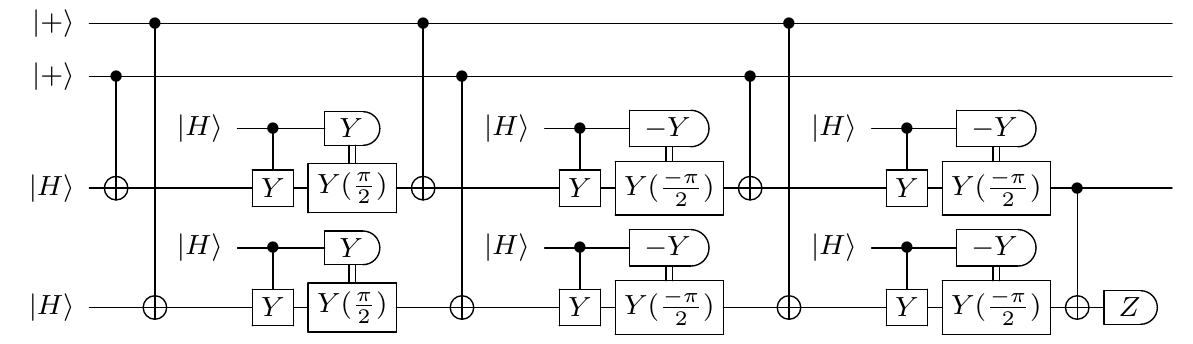}
\caption{$H$-to-Toffoli distillation circuit, expressed using exclusively Clifford gates and $\ket{H}$ states.  In total, this circuit uses $8$ $\ket{H}$ states and $36$ locations to distill a single Toffoli-state output.  For the purpose of counting locations, measurements and unitary one-qubit Clifford gates are ignored.
\label{fig:overheadHToToffoliDistillation}}
\end{figure*}

\begin{figure*}
\includegraphics{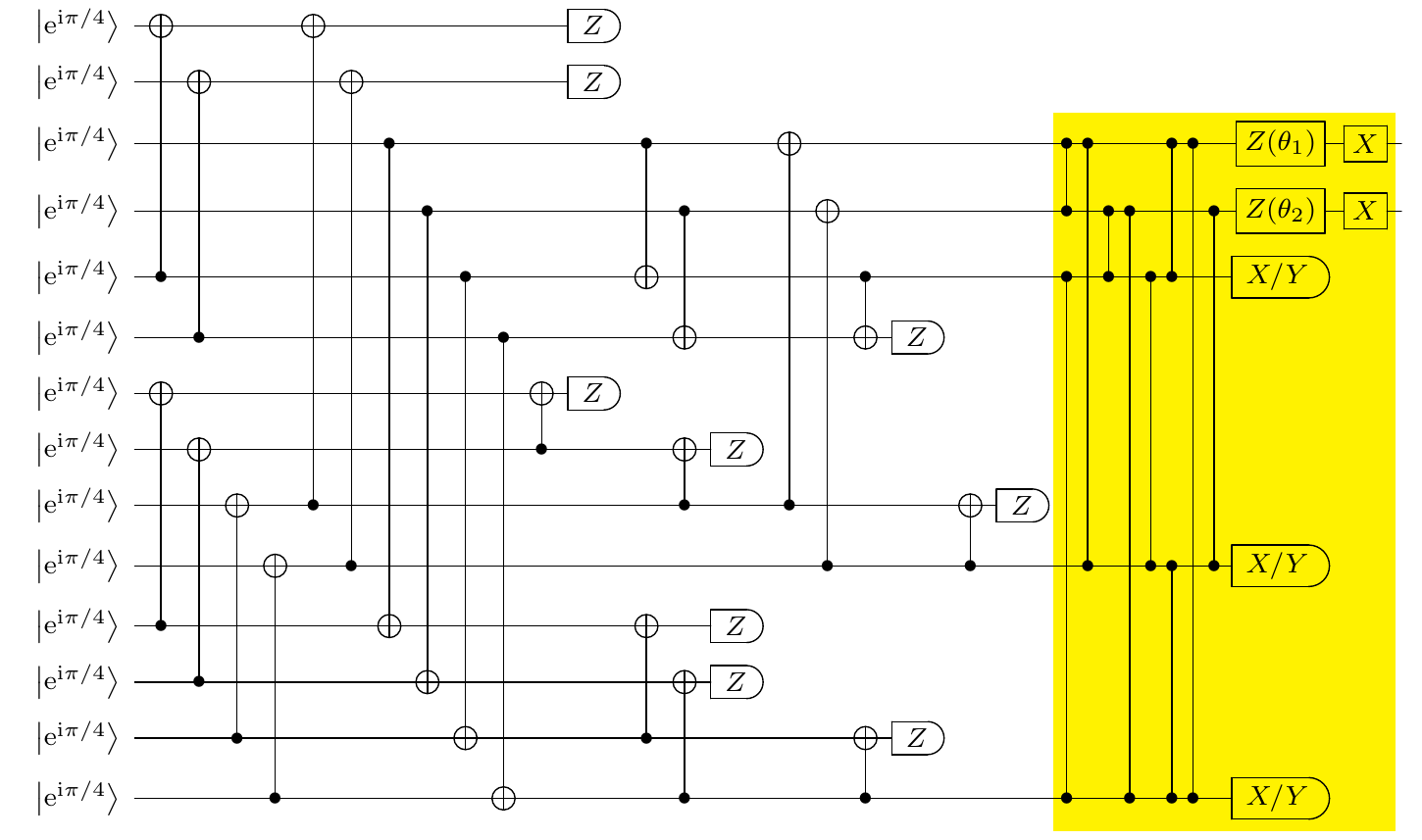}
\caption{(Color online) Compacted circuit for the $14$-to-$2$ distillation routine~\cite{Bravyi12} expressed in terms of Clifford gates and $\ket{\e^{\i \pi/4}}$ states.  The $\protect\CZ$ gates in the highlighted region act on every pair of unmeasured qubits.  Depending on the $Z$-measurement outcomes, some subset of the highlighted gates are implemented.  The angles $\theta_1$ and $\theta_2$ are multiples of $\pi/2$, where the multiple is likewise dependent on the $Z$-measurement outcomes.  The output is discarded whenever the outcome of any of the remaining measurements is non-trivial.  Using standard circuit identities, it can be shown that any distillation routine of the sort proposed by Bravyi and Haah in Ref.~\cite{Bravyi12} can be expressed in this form, though efficiency-wise it is not always desirable.  In total, this circuit uses $14$ $\ket{\e^{\i \pi/4}}$ states and at most $78$ locations to distill $2$ improved copies of $\ket{\e^{\i \pi/4}}$.  Conditional on success, the marginal probability of error is reduced from $p$ to roughly $7 p^2$.  For the purpose of counting locations, measurements and unitary one-qubit Clifford gates are ignored.  Note that $\ket{\e^{\i \pi/4}} = (\ket{0}+\e^{\i \pi/4}\ket{1})/\sqrt{2} = \e^{-\i \pi/8} H Z(\frac{-\pi}{2}) \ket{H}$ so the states $\ket{\e^{\i \pi/4}}$ and $\ket{H}$ are equivalent for the purpose of location counting.  \label{fig:overhead14to2Distillation}}
\end{figure*}

\bibliography{../citations}

\appendix*

\section{Location counting circuits\label{sec:locationCountingCircuits}}

\begin{sidewaysfigure}
\vspace{9.5cm}
\includegraphics{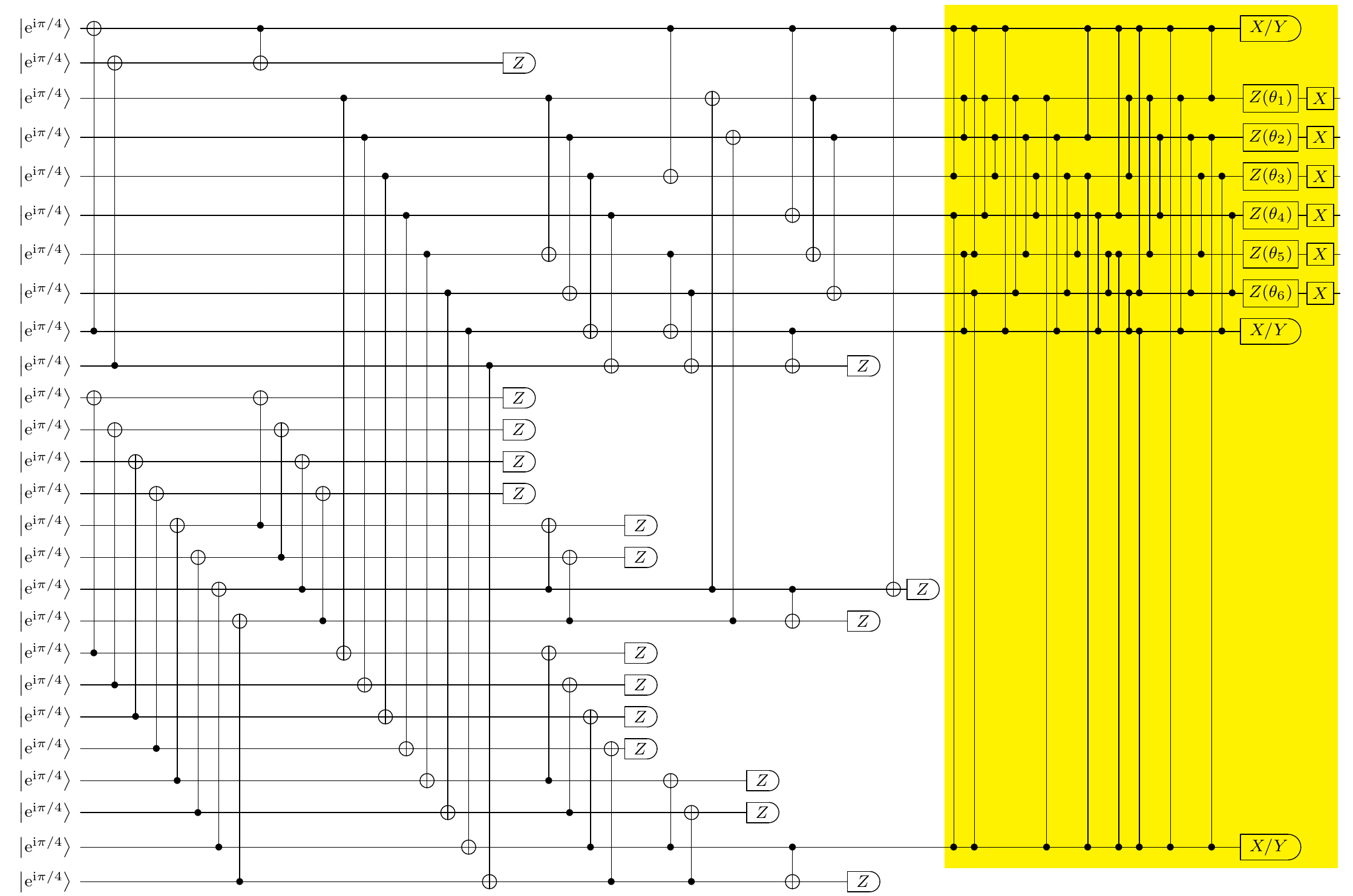}
\caption{(Color online) Compacted circuit for the $26$-to-$6$ distillation routine~\cite{Bravyi12} expressed in terms of Clifford gates and $\ket{\e^{\i \pi/4}} = (\ket{0}+\e^{\i \pi/4}\ket{1})/\sqrt{2}$ states. Depending on the $Z$-measurement outcomes, some subset of the highlighted gates are implemented.  The angles $\theta_i$ are multiples of $\pi/2$, where the multiple is likewise dependent on the $Z$-measurement outcomes.  The output is discarded whenever the outcome of any of the remaining measurements is non-trivial.  In total, this circuit uses $26$ $\ket{\e^{\i \pi/4}}$ states and at most $192$ locations to distill $6$ improved copies of $\ket{\e^{\i \pi/4}}$.  Conditional on success, the marginal probability of error is reduced from $p$ to roughly $76 p^2$.  For the purpose of counting locations, measurements and unitary one-qubit Clifford gates are ignored.\label{fig:overhead26to6Distillation}}
\end{sidewaysfigure}

\begin{figure*}
\includegraphics[clip = true, trim = .38cm 0cm .2cm 0cm]{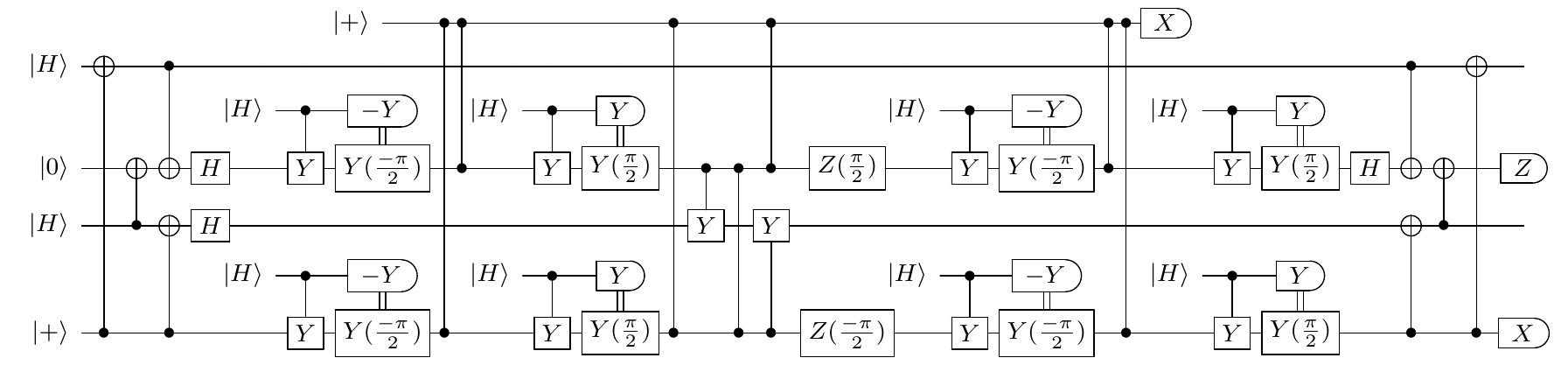}
\caption{Circuit for the $10$-to-$2$ magic-state distillation routine~\cite{Meier12} expressed in terms of Clifford gates and $\ket{H}$ states.  The output is discarded whenever a non-trivial measurement outcome is obtained.  In total, this circuit uses $10$ $\ket{H}$ states and $80$ locations to distill $2$ improved copies of $\ket{H}$.  Conditional on success, the marginal probability of error is reduced from $p$ to roughly $9 p^2$.  For the purpose of counting locations, measurements and unitary one-qubit Clifford gates are ignored.  \label{fig:overhead10to2Distillation}}
\end{figure*}

The location costs quoted for distillation routines in this paper were obtained using the circuits shown in Figs.~\ref{fig:overheadHToToffoliDistillation}, \ref{fig:overhead14to2Distillation}, \ref{fig:overhead26to6Distillation}, and \ref{fig:overhead10to2Distillation}.

\end{document}